\documentclass{llncs}
\input{psfig.sty}
\usepackage {graphicx}
\usepackage{amsmath}

\begin{document}
\pagestyle{headings}

\mainmatter

\title{Solving the subset-sum problem with a light-based device}

\author{Mihai Oltean, Oana Muntean}
\institute{Department of Computer Science,\\
Faculty of Mathematics and Computer Science,\\
Babe\c s-Bolyai University, Kog\u alniceanu 1,\\
Cluj-Napoca, 400084, Romania.\\
\email{moltean@cs.ubbcluj.ro}\\
\email{www.cs.ubbcluj.ro/$\sim$moltean}
}

\maketitle

\begin{abstract}

We propose a special computational device which uses light rays for solving the  subset-sum problem. The device has a graph-like representation and the light is traversing it by following the routes given by the connections between nodes. The nodes are connected by arcs in a special way which lets us to generate all possible subsets of the given set. To each arc we assign either a number from the given set or a predefined constant. When the light is passing through an arc it is delayed by the amount of time indicated by the number placed in that arc. At the destination node we will check if there is a ray whose total delay is equal to the target value of the subset sum problem (plus some constants). 

\end{abstract}

\section{Introduction}

Unconventional computing means computing by using new or unusual methods. Special properties of matter are usually exploited in an unconventional manner. Several unconventional techniques and approaches for attacking difficult problems have been investigated so far: DNA computing \cite{adleman}, Quantum computing \cite{Shor}, Soap bubbles \cite{Aaronson,Bringsjord}, Gear-based computers \cite{Vergis}, Adiabatic algorithms \cite{Kieu} etc.

Using light, instead of electric power, for performing computations is an exciting idea whose applications can be already seen on the market. This choice is motivated by the increasing number of real-world problems where the light-based devices could perform better than electric-based counterparts. Good examples in this direction can be found in the field of Optical Character Recognition \cite{wiki}. Another interesting example is the $n$-point discrete Fourier transform which can be performed in unit time by special light-based devices \cite{goodman,reif}.

In this paper we suggest a new way of performing computations by using some properties of light. The idea is used within a special device for solving the subset-sum problem. The problem asks to find if there is a subset of given set $A$ whose sum is $B$.

The device, which is very simple, has a graph-like structure. The nodes are connected by arcs in such way all possible subsets of $A$ are generated. To each arc we assign either a number from the given set $A$, or a predefined constant. The length of an arc is directly proportional to the number assigned to it. 

Initially a light ray is sent to the start node. In each node the light is divided into 2 subrays. Each arc delays the ray by an amount of time equal to the number assigned to it. At the destination node we will check if there is a ray arriving in the destination node at the moment equal to the target value of $B$ (plus some constants introduced by the system).

The paper is organized as follows: Related work in the field of optical computing is briefly overviewed in section \ref{related}. The subset-sum problem is described in section \ref{subsetsum}. The proposed device is presented in section \ref{proposed}. The way in which the proposed device works is given in section \ref{howorks}. A list of components required by the proposed device is given in section \ref{hard}. Complexity is computed in section \ref{complexity}. Weaknesses of our device are discussed in section \ref{dif}. Section \ref{psize} gives a rough approximation for the size of the instances that can be solved by our device. Suggestions for improving the device are given in section \ref{speed_reduce}. Further work directions are suggested in section \ref{further}.

\section{Related work}
\label{related}

Most of the major computational devices today are using electric power in order to perform useful computations. 

Another idea is to use light instead of electrical power. It is hoped that optical computing could advance computer architecture and can improve the speed of data input and output by several orders of magnitude \cite{Feitelson}.

Many theoretical and practical light-based devices have been proposed for dealing with various problems. Optical computation has some advantages, one of them being the fact that it can perform some operations faster than conventional devices. An example is the $n$-point discrete Fourier transform computation which can be performed, optically, in only unit time \cite{goodman,reif}. Based on that, a solution to the subset sum problem can be obtained by discrete convolution. The idea is that the convolution of 2 functions is the same as the product of their frequencies representation \cite{idrone}.

The quest for the light-based computers was started in 1929 by G. Tauschek who has obtained a patent on Optical Character Recognition (OCR) in Germany. Next step was made by Handel who obtained a patent on OCR. Those devices were mechanical and used templates for matching the characters. A photodetector was placed so that when the template and the character to be recognized were lined up for an exact match, and a light was directed towards it, no light would reach the photodetector \cite{wiki}. Since then, the field of OCR has grown steadily and recently has become an umbrella for multiple pattern recognition techniques (including Digital Character Recognition).

An important practical step was made by Intel researchers who have developed the first continuous wave all-silicon laser using a physical property called the Raman Effect \cite{Faist,Paniccia,rong1,rong2}. The device could lead to such practical applications as optical amplifiers, lasers, wavelength converters, and new kinds of lossless optical devices.

\begin{figure}[htbp]
\centerline{\includegraphics[width=3.83in,height=5.21in]{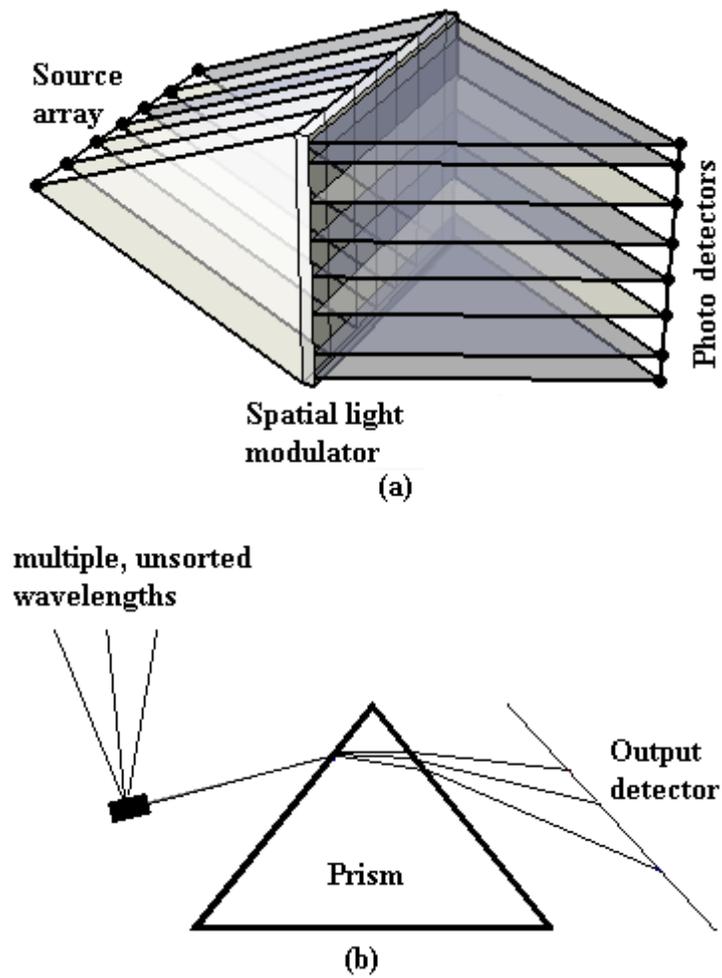}}
\caption{(a) A sketch of the Lenslet device used for performing vector-matrix multiplications. (b) Schematic view of the Rainbow Sort.}
\label{fig_big}
\end{figure}

Another solution comes from Lenslet \cite{lenslet} which has created a very fast processor for vector-matrix multiplications (see Figure \ref{fig_big} (a)).  This processor can perform up to 8000 Giga Multiple-Accumulate instructions per second. Lenslet technology has already been applied to data analysis using $k-$mean algorithm and video compression.

A recent paper \cite{Schultes} introduces the idea of sorting by using some properties of light. The method called Rainbow Sort is based on the physical concepts of refraction and dispersion. It is inspired by the observation that light that traverses
a prism is sorted by wavelength (see Figure \ref{fig_big} (b)). For implementing the Rainbow Sort one need to perform the following steps:

\begin{itemize}

\item{encode multiple wavelengths (representing the numbers to be sorted) into a light ray,}

\item{send the ray through a prism which will split the ray into $n$ monochromatic rays that are sorted by wavelength,}

\item{read the output by using a special detector that receives the incoming rays.}

\end{itemize}

A stable version of the Rainbow Sort is proposed in \cite{Murphy}.

Naughton (et al.) proposed and investigated \cite{Naughton,woods} a model called the continuous space machine which operates in discrete time-steps over a number of two-dimensional complex-valued images of constant size and arbitrary spatial resolution. The (constant time) operations on images include Fourier transformation, multiplication, addition, thresholding, copying and scaling.

A system which solves the Hamiltonian path problem (HPP) \cite{garey} by using light and its properties has been proposed in \cite{oltean_uc,oltean_nc}. The device has the same structure as the graph
where the solution is to be found. The light is delayed within nodes, whereas the delays introduced by arcs are constants. Because the problem asks that each node has to be visited exactly once, a special delaying system was designed. At the destination node we will search for a ray which has visited each node exactly once. This is very easy due to the special properties of the delaying system.

\section{The subset-sum problem}
\label{subsetsum}

The description of the subset-sum problem \cite{cormen1,garey} is the following:

Given a set of positive numbers $A = \{a_1, a_2, ..., a_n\}$ and another positive number $B$. Is there a subset of $A$ whose sum equals $B$?

We focus our attention on the YES / NO decision problem. We are not interested in finding the subset generating the solution. Actually we are interested to find only if such subset does exist.

The subset-sum problem arises in many real-word applications \cite{gilmore}.

The problem belongs to the class of NP-complete problems \cite{garey}. No polynomial time algorithm is known for it. However, a pseudo-polynomial time algorithm does exist for this problem \cite{garey}. The complexity of this algorithm is bounded by both $n$ and $B$. The algorithm requires $O(n * B)$ storage space.

\section{The proposed device}
\label{proposed}

This section deeply describes the proposed system. Section \ref{useful} describes the properties of light which are useful for our device. Section \ref{operations} introduces the operations performed by the components of our device. Basic ideas behind our concept are given in section \ref{basic}. Some examples on how the system works are given in section \ref{howorks}.

\subsection{Useful properties of light}
\label{useful}

Our idea is based on two properties of light:

\begin{itemize}

\item{The speed of light has a limit. The value of the limit is not very important at this stage of explanation. The speed will become important when we will try to measure the moment when rays arrive at the destination node (see section \ref{precision}). What is important now is the fact that we can delay the ray by forcing it to pass through an optical fiber cable of a certain length.}

\item{The ray can be easily divided into multiple rays of smaller intensity/power. Beam-splitters are used for this operation \cite{Agrawal,Feitelson}.}

\end{itemize}

\subsection{Operations performed within our device}
\label{operations}

The proposed device has a graph like structure. Generally speaking one operation is performed when a ray passes through a node and one operation is performed when a ray passes through an edge.

\begin{itemize}

\item{When passing through an arc the light ray is delayed by the amount of time assigned to that arc.}

\item{When the ray is passing through a node it is divided into a number of rays equal to the external degree of that node. Each obtained ray is directed toward one of the nodes connected to the current node.}

\end{itemize}

\subsection{The device}
\label{basic}

The first idea for our device was that numbers from the given set $A$ represent the delays induced to the signals (light) that passes through our device. For instance, if numbers $a_1$, $a_3$ and $a_7$ generate the expected subset, then the total delay of the signal should be $a_1 + a_3 + a_7$. If using light we can easily induce some delays by forcing the ray to pass through an optical cable of given length. 

\begin{figure}[htbp]
\centerline{\includegraphics[width=4.52in,height=1.21in]{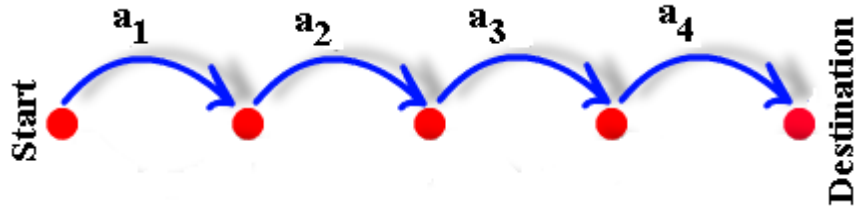}}
\caption{First version of our device. Each arc delays the ray by the amount of time written on it. Note that this device is not complete because it cannot generate all possible subsets of $A$.}
\label{fig_device1}
\end{figure}

This is why we have designed our device as a directed graph. Arcs, which are implemented by using optical cables, are labeled with numbers from the given set $A$. Each number is assigned to exactly one arc and there are no two arcs having assigned the same number. There are $n + 1$ nodes connected by $n$ arcs. At this moment of explanation we have a linear graph as the one shown in Figure \ref{fig_device1}. 

However, this is not enough because we also need a mechanism for skipping an arc. Only in this way we may generate all possible subsets of $A$. 

\begin{figure}[htbp]
\centerline{\includegraphics[width=4.51in,height=1.33in]{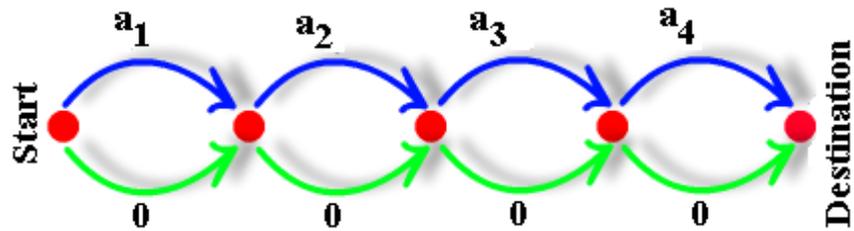}}
\caption{Second version of our device. Each subset of $A$ is generated, but this device cannot be implemented in practice because we cannot have cables of length 0.}
\label{fig_device2}
\end{figure}

A possible way for achieving this is to add an extra arc (of length 0) between any pair of consecutive nodes. Such device is depicted in Figure \ref{fig_device2}. A light ray sent to start node will have the possibility to either traverse a given arc (from the upper part of figure) or to skip it (by traversing the arc of length 0 from the bottom of figure).

In each node (but the last one) we place a beam-splitter which will split a ray into 2 subrays of smaller intensity. 

The device will generate all possible subsets of $A$. Each subset will delay one of the ray by an amount of time equal to the sum of the lengths of the arcs in that path.

There is a problem here: even if theoretically we could have arcs of length 0, we cannot have cables of length 0 in practice. For avoiding this problem we have multiple solutions. The first one was to use very short cables (let's say of length $\epsilon$) for arcs which are supposed to have length 0. However, there is another problem here: we could obtain for instance the sum $B$ written as $B = a_1 + 3*\epsilon$. Even if there is no subset of sum $B$, still there will be possible to have a signal at moment $B$ due to the situation presented above.

\begin{figure}[htbp]
\centerline{\includegraphics[width=4.52in,height=1.34in]{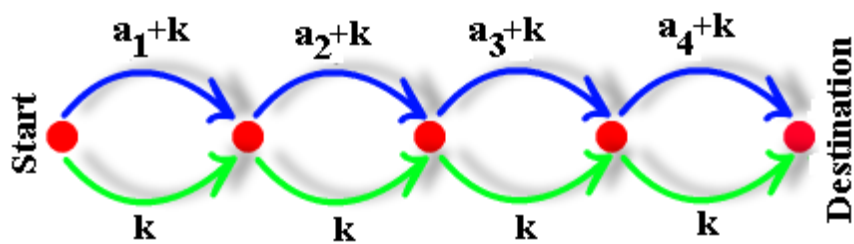}}
\caption{A schematic representation of the device used for solving an instance with 4 numbers. On each arc we have depicted its length. There are $n$ cables of length $k$ and $n$ cables of length $a_i + k$ ($1 \le i \le n$). This device does generate all possible subsets of $A$ and it can be implemented in practice}
\label{fig_device3}
\end{figure}

For avoiding this situation we have added a constant $k$ to the length of each cable. The schematic view of this device is depicted in Figure \ref{fig_device3}.

We can see that each path from $Start$ to $Destination$ contains exactly $n$ time value $k$. Thus, at the destination we will not wait anymore at moment $B$. Instead we will wait for a solution at moment $B+n*k$ since all subsets will have the constant $n*k$ added.

The device will generate all possible subsets of $A$. The good part is that we do not have to check all $2^n$ possible solutions. We will only have to check if there is a ray arriving at moment $B+n*k$ in the destination node. The signals generated by all other subsets are ignored and not recorded in any way.

\subsection{How the system works}
\label{howorks}

\begin{figure}[htbp]
\centerline{\includegraphics[width=4.79in,height=4.45in]{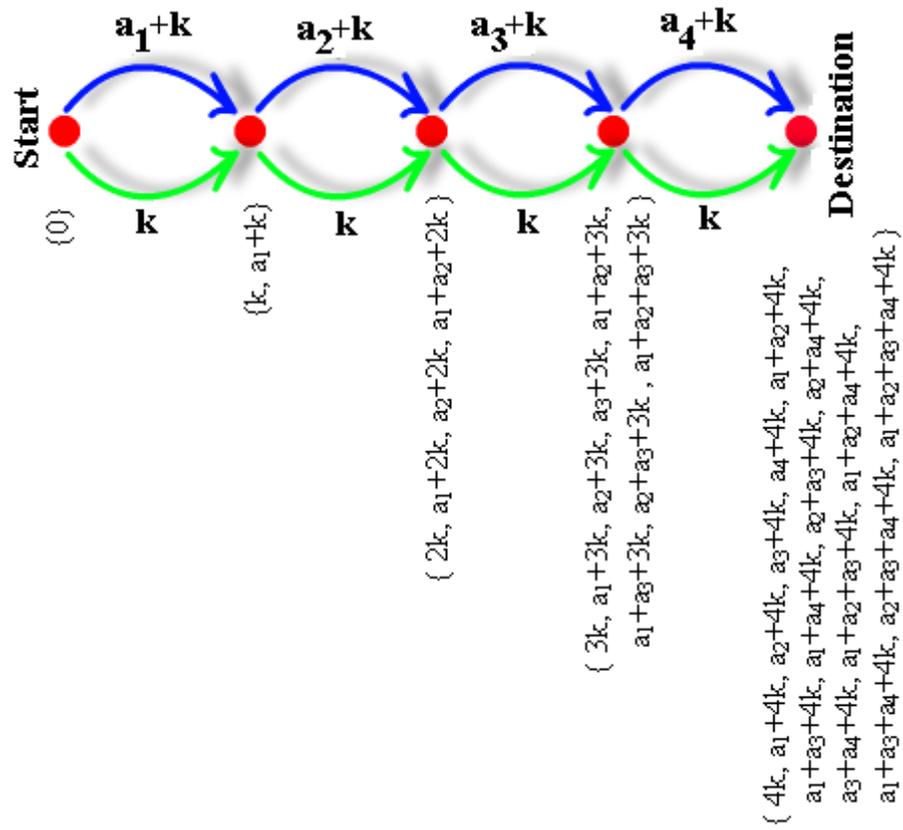}}
\caption{The moments when different rays arrive in nodes. The moments are represented as sets because they might not be distinct}
\label{fig_device4}
\end{figure}

In the graph depicted in Figure \ref{fig_device4} the light will enter in $Start$ node. It will be divided into 2 subrays of smaller intensity. These 2 rays will arrive into the second node at moments $a_1+k$ and $k$. Each of them will be divided into 2 subrays which will arrive in the $3^{rd}$ node at moments $2*k, a_1+2*k, a_2+2*k, a_1+a_2+2*k$. These rays will arrive at no more than 4 different moments. 

In the destination node we will have $2^n$ rays arriving at no more than $2^n$ different moments. The ray arriving at moment $n*k$ means the empty set. The ray arriving at moment $a_1+a_2+...+a_n + n*k$ represents the full set. If there is a ray arriving at moment $B+n*k$ means that there is a subset of $A$ of sum $B$.

If there are 2 rays arriving at the same moment in the $Destination$ it simply means that there are multiple subsets which have the same sum. This is not a problem for us because we want to answer the YES/NO decision problem (see section \ref{subsetsum}). We are not interested at this moment which is the subset generating the solution.

Because we are working with continuous signal we cannot expect to have discrete output at the destination node. This means that rays arrival is notified by fluctuations in the intensity of the light. These fluctuations will be transformed, by a photodiode, in fluctuations of the electric power which will be easily read by an oscilloscope.

\section{Physical implementation}
\label{hard}

\begin{figure}[htbp]
\centerline{\includegraphics[width=1.95in,height=1.76in]{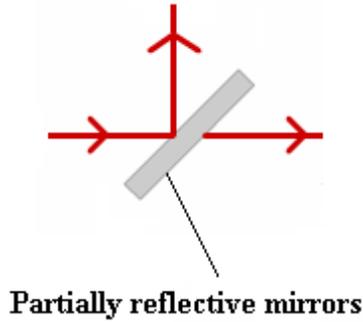}}
\caption{The way in which a ray can be split into 2 sub-rays by using a beam-splitter.}
\label{beam_splitter}
\end{figure}

For implementing the proposed device we need the following components:

\begin{itemize}

\item{a source of light (laser),}

\item{Several beam-splitters for dividing light rays into 2 subrays. A standard beam-splitter is designed using a half-silvered mirror (see Figure \ref{beam_splitter}),}

\item{A high speed photodiode for converting light rays into electrical power. The photodiode is placed in the destination node,}

\item{A tool for detecting fluctuations in the intensity of electric power generated by the photodiode (oscilloscope),}

\item{A set of optical fiber cables having lengths equals to the numbers in the given set $A$ (plus constant $k$) and another set of $n$ cables having fixed length $k$. These cables are used for connecting nodes. }

\end{itemize}

\section{Complexity}
\label{complexity}

The time required to build the device has $O(n*B)$ complexity. We assume that all cables are shorter than $B$, otherwise they cannot participate to the final solution.

Because the ray encoding the solution takes $B+n*k$ time to reach the destination node we may say that the complexity is $O(B+n)$.

The intensity of the signal decreases exponentially with the number of nodes. This is why the required power is proportional to $2^n$.

\section{Analysis of the proposed device}
\label{dif}

This section investigates some of the problems of the proposed device and some ways to deal with them. Section \ref{precision} computes the precision of solution representation. The size of the instances that can be solved by our device having a limited length for cables is computed in section \ref{psize}. Section \ref{amplify} shows how to handle the exponential decrease of power. Several problems that might be encountered during the physical implementation are discussed in section \ref{tech}. Section \ref{speed_reduce} shows how to improve the device by reducing the speed of light.

\subsection{Precision}
\label{precision}

A problem is that we cannot measure the moment $B+n*k$ exactly. We can do this measurement only with a given precision which depends on the tools involved in the experiments. Actually it will depend on the response time of the photodiode and the rise time of the oscilloscope.

The rise-time of the best oscilloscope available on the market is in the range of picoseconds ($10^{-12}$ seconds). This means that if a signal arrives at the destination in the interval $[B+n*k-10^{-12}, B+n*k+10^{-12}]$ we cannot be perfectly sure that we have a correct subset or another one which does not have the wanted property. This problem can be avoided if all cables are long enough. In what follows we will try to compute the length of the cables.

We know that the speed of light is $3 \cdot 10^{8} m/s$. Based on that we can easily compute the minimal cable length that should be traversed by the ray in order to be delayed with $10^{-12}$ seconds. This is obviously 0.0003 meters and it was obtained from equation:

\[
distance = speed * time
\]

This value is the minimal delay that should be introduced by an arc. More than that, all lengths must be integer multiples of 0.0003. We cannot allow to have cables whose lengths can be written as $p*0.0003 + q$, where $p$ is an integer and $q$ is a positive real number less than 0.0003 because by combining this kind of numbers we can have a signal in the above mentioned interval and that signal does not encode a subset whose sum is the expected one.

Once we have the length for the minimal delay it is quite easy to compute the length of the other cables that are used in order to induce a certain delay. First of all we have to multiply / divide all given numbers with such factor that the less significant digit (greater than 0) to be on the first position before the decimal place. For instance if we have the set $A = \{0.001, 4\}$ we will multiply both numbers by 1000. If we have the set $A = \{100, 2000\}$ we have to divide both numbers by 100. After this operation we will multiply the obtained numbers by 0.0003 factor.

This will ensure that if a signal will arrive in the interval $[B+n*k-10^{-12}, B+n*k+10^{-12}]$ we can be sure that it encodes the sum $B+n*k$.

\subsection{Problem size}
\label{psize}

We are also interested to find the size of the instances that can be solved by our device. Regarding the cardinal of $A$ we cannot make too many approximations because it actually depends on the available power and on the sensitivity of the measurement tools.

However, having available a limited length (lets say 3 kilometers) for each cable, we can compute the maximal value for the numbers that can appear in $A$. We know that each number is less or equal to $B$. This is why we want to see how large $B$ can be.  

Without reducing generality we may assume that all numbers are positive integers. We know that the shortest delay possible is 0.0003 meters (see section \ref{precision}). Having a cable of 3 kilometers we may encode numbers less than $10^7$.

Longer cables may also be available. Take for instance the optical cables linking the cities in a given country. We may easily find cables having 300 km. In this case we may work with numbers smaller than $10^9$. This is a little bit smaller than the largest integer value represented over 32 bits.

\subsection{Power decrease}
\label{amplify}

Beam splitters are used in our approach for dividing a ray in two subrays. Because of that, the intensity of the signal is decreasing. In the worst case we have an exponential decrease of the intensity. For instance, in a graph with $n$ nodes (destination node is not counted because there is no split there), each signal is divided (within each node) into 2 signals. Roughly speaking, the intensity of the signal will decrease $2^n$ times.

This means that, at the destination node, we have to be able to detect very small fluctuations in the intensity of the signal. For this purpose we can use a photomultiplier \cite{Flyckt} which is an extremely sensitive detector of light in the ultraviolet, visible and near infrared range. This detector multiplies the signal produced by incident light by as much as $10^8$, from which even single photons can be detected.

Also note that this difficulty is not specific to our system only. Other major unconventional computation paradigms, trying to solve NP-complete problems share the same fate. For instance, a quantity of DNA equal to the mass of Earth is required to solve Hamiltonian Path Problem instances of 200 cities using DNA computers \cite{Hartmanis}.

\subsection{Technical difficulties}
\label{tech}

There are many technical challenges that must be solved when implementing the proposed device. Some of them are:

\begin{itemize}

\item{Cutting the optic fibers to an exact length with high precision. Failing to accomplish this task can lead to errors in detecting if there was a fluctuation in the intensity at moment $B+n*k$,}

\item{Finding a high precision oscilloscope. This is an essential step for measuring the moment $B+n*k$ with high precision (see section \ref{precision}).}

\end{itemize}

\subsection{Improving the device}
\label{speed_reduce}

The speed of the light in optic fibers is an important parameter in our device. The problem is that the light is too fast for our measurement tools. We have either to increase the precision of our measurement tools or to decrease the speed of light.

It is known that the speed of light traversing a cable is significantly smaller than the speed of light in the void space. Commercially available cables have limit the speed of the ray wave up to $60\%$ from the original speed of light. This means that we can obtain the same delay by using a shorter cable.

However, this method for reducing the speed of light is not enough for our purpose. The order of magnitude is still the same. This is why we have the search for other methods for reducing that speed. A very interesting solution was proposed in \cite{Hau} which is able to reduce the speed of light by 7 orders of magnitude and even to stop it \cite{Bajcsy,Liu}. In \cite{Bajcsy} they succeeded in completely halting light by directing it into a mass of hot rubidium gas, the atoms of which, behaved like tiny mirrors, due to an interference pattern in two control beams.

This could help our mechanism significantly. However, how to use this idea for our device is still an open question because of the complex equipment involved in those experiments \cite{Hau,Liu}.

By reducing the speed of light by 7 orders of magnitude we can reduce the size of the involved cables by a similar order (assuming that the precision of the measurement tools is still the same). This will help us to solve larger instances of the problem.

\section{Conclusions and further work}
\label{further}

The way in which light can be used for performing useful computations has been suggested in this paper. The techniques are based on the massive parallelism of the light ray.

It has been shown the way in which a light-based device can be used for solving the subset-sum problem. 

Further work directions will be focused on:

\begin{itemize}

\item{implementing the proposed device,}

\item{cutting new cables each time when a new instance has to be solved is extremely inefficient. This is why finding a simple way to reuse the previously utilized cables is a priority for our system,}

\item{automate the entire process,}

\item{Our device cannot find the set of numbers representing the solution. It can only say if there is a subset or not. If there are multiple subsets we cannot distinguish them. However, the subset sum YES/NO decision problem is still a NP-complete problem \cite{garey}. We are currently investigating a way to store the order of nodes so that we can easily reconstruct the path,}

\item{finding other non-trivial problems which can be solved by using the proposed device,}

\item{finding other ways to introduce delays in the system. The current solution requires cables that are too long and too expensive,}

\item{using other type of signals instead of light. Possible candidates are electric power and sound.}

\end{itemize}

\end{document}